# Scanned Potential Microscopy of Edge and Bulk Currents in the Quantum Hall Regime

Kent L. McCormick, Michael T. Woodside, Mike Huang, Mingshaw Wu, and Paul L. McEuen
*Department of Physics, University of California and Materials Science Division, Lawrence Berkeley National Laboratory, Berkeley, California, 94720*

C. I. Duruoz and J. S. Harris Jr.
*Dept. of Electrical Engineering, Stanford University, Stanford, California, 94309*

Using an atomic force microscope as a local voltmeter, we measure the Hall voltage profile in a 2D electron gas in the quantum Hall (QH) regime. We observe a linear profile in the bulk of the sample in the transition regions between QH plateaus and a distinctly nonlinear profile on the plateaus. In addition, localized voltage drops are observed at the sample edges in the transition regions. We interpret these results in terms of theories of edge and bulk currents in the QH regime.

Since the discovery of the quantum Hall (QH) effect [1], the electrical characteristics of quantum Hall conductors (QHCs) have been intensely studied. The universal nature of the quantization in QHCs leads to a resistance that is independent of the microscopic properties of the sample. As a consequence, local properties of a QHC such as the current and voltage distributions within the sample are inaccessible with standard transport measurements. These local properties remain controversial, with some theories suggesting that the current flow and associated voltage drops are concentrated at the edges of the sample [2-5], while others predict a distribution extending throughout the bulk of the sample [6-8]. Attempts have been made to address these questions with local imaging techniques [9-13], transport measurements of the breakdown of the QH effect [14], inductive measurements [15], capacitance measurements [16], and internal voltage probes [17]. The results, while informative, have lacked sufficient spatial and/or energy resolution to determine unambiguously how the current is partitioned between edge and bulk channels and the shape of the associated voltage profile across the QHC.

To address these issues, we have used a scanned potential microscope to study the potential distribution in a QHC with submillivolt voltage and submicron spatial resolution. We find that at low magnetic fields the Hall voltage drop is linear, indicating uniform current flow throughout the sample. At high magnetic fields, large voltage drops are seen at the sample edges in the transition regions between QH plateaus, indicating that the current is concentrated near the edges. On the QH plateaus, the current is distributed throughout the bulk in a complex, non-uniform way. These results are in good agreement with existing results for edge and bulk transport in the QH regime.

The samples consist of 10-20 μm wide Hall bars defined by wet etching of a GaAs/AlGaAs heterostructure with a 2D electron gas lying 77 nm beneath the surface. Results will be presented on two samples with mobilities of 8 and 70 m$^2$/V-s and densities of 2.8 x 10$^{15}$ and 2.6 x 10$^{15}$ m$^{-2}$. All measurements were performed at temperatures between 0.7 and 1.0 K. The samples were also characterized by standard transport measurements.

We measure the local voltage with a low-temperature atomic force microscope (AFM) operating in non-contact mode [18]. As shown schematically in Fig. 1, an AC potential $V_0$ is applied to the contacts of the sample, producing inside the sample the AC potential $V(x,y)$ whose spatial distribution is to be measured. The local sample potential $V(x,y)$ interacts electrostatically with the sharp, metallized AFM tip, deflecting the AFM cantilever with a force that can be simply modelled as [19]:

$$F_{ac}(x,y) = \frac{dC}{dz}(V_{dc} + \Phi)V(x,y).$$

C is the tip-sample capacitance, z is the tip-sample separation, $V_{dc}$ is the DC voltage applied between tip and sample, and $\Phi$ is the contact potential difference between the tip and the sample materials. The force on the tip, and hence the deflection of the cantilever, is thus directly proportional to $V(x,y)$, the local electrostatic potential in the sample [20].

Under typical operating conditions, z = 50 nm, dC/dz ~ 5x10$^{-11}$ F/m [21], ($V_{dc}$ + $\Phi$) ~ 0.4V, and $V_0$ ~ 1 mV. The resulting cantilever deflection of ~ 2 nm is detected with a piezoresistive sensor [22]. To enhance the force sensitivity, the frequency of the driving voltage $V_0$ is maintained at the resonant frequency of the cantilever, ~120 kHz. Under these conditions, the voltage sensitivity is about 10 μV/Hz$^{1/2}$, limited by thermal fluctuations in the cantilever [23]. The spatial resolution is approximately 200 nm. The DC voltage on the tip perturbs the carrier density in the sample by less than 10% [24], as estimated from scanned gate measurements [25], but changing the density perturbation from about 5% to 25% does not qualitatively affect the observations reported below. Note also that height and contact potential fluctuations [26] cause local variations in the signal strength. We account for these by normalizing the measured signal with a simultaneously-measured reference signal whereby a uniform voltage is applied to all contacts on the sample. Further details of the design and operation are discussed elsewhere [24].

We first study the magnetic field dependence of the Hall voltage profile V(x) perpendicular to the current flow by repeatedly scanning across the width of the Hall bar while changing the magnetic field. At high fields, the Hall voltage profile can be used to infer the current density $j_y(x)$ with the relation:

$$j_y(x) \approx -\sigma_{xy} \frac{\partial V(x)}{\partial x}.$$

The results are displayed in Fig. 2. At B= 0 T, the voltage is uniform across the width of the sample. As B is increased from 0, a Hall voltage develops, with a linear spatial profile at low fields. At higher fields a cyclic pattern emerges that is periodic in $B^{-1}$. This pattern is commensurate with the QH plateaus observed in standard transport measurements on the same sample (not shown).

Fig. 3 shows the evolution of V(x) through the cycle around ν = 6 in greater detail. At fields well below the QH plateau, the voltage profile is linear (Fig. 3(a)). As the ν = 6 plateau is approached, the potential in the bulk remains linear but flattens out, and sharp drops develop at the edges (Figs. 3(b),(c)). These edge drops are a generic feature in the transition regions, occuring for 2n < ν < 2n+1 (integer n) in this sample. Fig. 4 shows an expanded view of the voltage peaks near the edge of the Hall bar, at three different filling factors. The width of the edge feature decreases with increasing B, changing from approximately 1 μm at ν = 6.5 to 300 nm at ν = 2.5.

Very different behavior is observed on the quantum Hall plateaus (Figs. 3(d)-(g)). The voltage profile becomes quite complicated and non-linear, developing significant gradients within the bulk of the sample that evolve rapidly with B across the plateau. These voltage gradients exist in both the x and y directions. In Fig. 5, the local potential V(x,y) on the $\nu = 4$ QH plateau of the high-mobility sample is displayed. A sharp drop in the Hall voltage V(x) in the bulk of the sample whose location depends upon y is particularly noticeable. Similar results were obtained for the low-mobility sample, but the fluctuations of the local voltage with position were smaller.

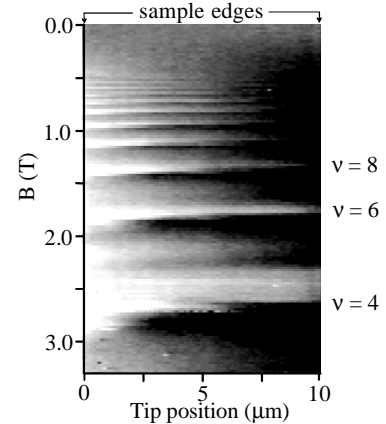

FIG. 2. Magnetic field dependence of Hall voltage profile across a 10 μm-wide Hall bar on the low-mobility sample. White indicates high voltage, black low voltage. Near 0.5 T, structure periodic in $B^{-1}$ and commensurate with even filling factors appears (the ν = 8, 6, and 4 QH plateaus are marked). Similar results are seen on the high-mobility sample.

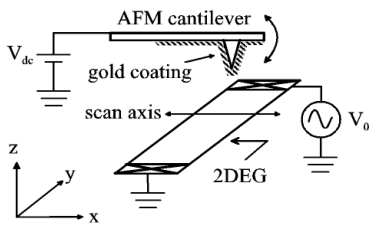

FIG. 1. Schematic diagram of measurement. Applying $V_0$ to one of the contacts creates a potential V(x,y) inside the sample which interacts electrostatically with the metallized AFM tip positioned 50 nm above the sample, causing the AFM cantilever to vibrate by an amount proportional to V(x,y). $V_0$ is kept at the resonant frequency of the cantilever by a self-resonant positive-feedback loop in order to maximize the sensitivity.

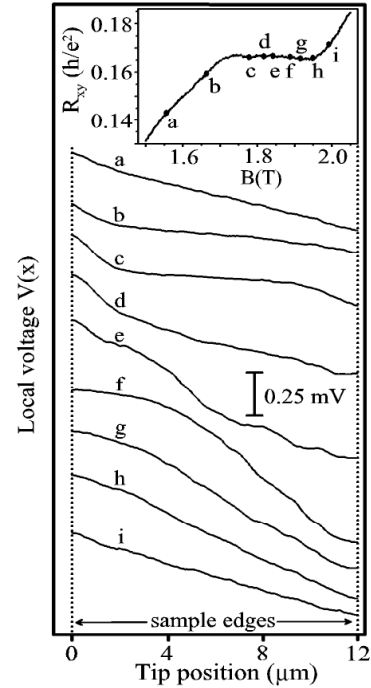

FIG. 3. Voltage profiles across a 12 μm-wide Hall bar on the low-mobility sample near the ν = 6 QH plateau (traces offset for clarity, voltage scale as shown). The B-fields at which the profiles were measured and the ν = 6 plateau measured by transport are shown in the inset. (a) Well below the plateau, the Hall profile is linear. (b,c) As the plateau is approached, the bulk retains a linear profile but decouples from the edges. (d-g) Well inside the plateau, the profile in the bulk becomes complicated and changes rapidly with field. (h,i) A linear profile returns upon leaving the plateau.

As B is increased further, leaving the QH plateau, the Hall voltage V(x) regains its linear profile (Figs. 3(h),(i)). The whole cycle described above repeats through the next QH transition and plateau. This cyclic evolution of the local voltage profile with magnetic field is generic and reproducible, with the same qualitative behavior seen at all filling factors studied ($\nu=10$ to $\nu=2$) at different locations in both high- and low-mobility samples. Only the details of the voltage gradients in the bulk on the QH plateaus are variable.

All of these observations can be understood within a relatively simple framework for transport in QHCs. At low magnetic fields, the 2DEG is metallic and is characterized by a uniform local conductivity. Under these circumstances, a linear voltage drop is expected [27]. The measurements at low field agree with this result.

In the QH regime, we distinguish between the behavior on the plateaus and that in the transition regions between them. In the transition regions, the bulk of the sample is conducting due to the existence of extended states at the Fermi energy $E_F$. As in the low-field case, there is a uniform local conductivity in the bulk, and a linear Hall voltage profile is expected. The situation at the edges of the sample is expected to be more complex, however, due to the existence of non-equilibrium edge states [5,28,29]. These states occur at the boundary of the sample, where the Landau levels (LL) that are occupied in the bulk cross $E_F$. Previous work has shown that these edge states can have any degree of coupling to the bulk state [28,29], and hence can be at a different electrochemical potential than the bulk state. If the edge and bulk states equilibrate, a linear voltage drop throughout the sample is expected. If they are out of equilibrium, significant voltage drops at the edges should be present.

Our observations indicate that, just after a QH plateau, the bulk and edges are well equilibrated and the Hall profile is linear throughout the sample (Figs. 3(a),(h),(i)). As the field is increased through the transition region, the local voltage drop that develops at the edges indicates disequilibration. The width of these edge features increases as the number of edge states present increases: the single (spin-degenerate) edge state at $\nu = 2.5$ has a width of 0.3 μm, very close to the resolution of our AFM, while at $\nu = 4.5$ the two states together have a width of 0.6 μm, and at $\nu = 6.5$ the three states together have a width of 1 μm. These sizes are consistent with previous estimates of the edge state widths from capacitance measurements [16] and theoretical models [5]. Since the magnitude of the voltage drops at the edges is large relative to the drop across the bulk (Fig. 3(c)), most of the Hall current flows at the edges for the filling factors shown in Fig. 4. At present, we are unable to resolve the internal structure expected due to the alternation of compressible and incompressible strips at the higher filling factors.

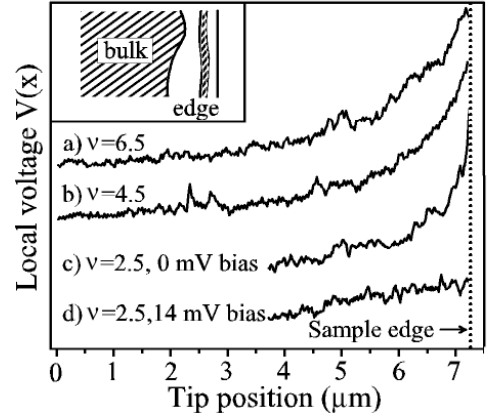

FIG. 4. Voltage profile at the edge of a 14 μm-wide Hall bar on the low-mobility sample (traces offset for clarity). (a-c) Sharp potential drops at the sample edge are seen at various filling factors where non-equilibrium edge states exist (see text). These arise because the edge states are at a different potential than the bulk (inset). The width of the edge signals at half-max is about 0.3 μm at $\nu = 2.5$, 0.6 μm at $\nu = 4.5$, and 1 μm at $\nu = 6.5$. (d) The edge signal disappears when a DC bias is applied, causing the bulk and edge states to equilibrate.

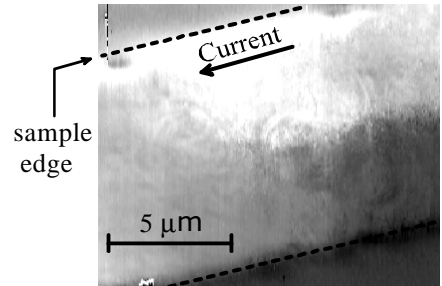

FIG. 5. Area scan of voltage on a Hall bar on the high-mobility sample at $\nu = 4$. White indicates high voltage, black low voltage. The position of a sharp voltage drop in the bulk (shown as an abrupt change in shade) varies considerably along the Hall bar.

This interpretation of edge states out of equilibrium with the bulk is supported by additional measurements. The influence of non-equilibrium edge states on standard transport measurements can be used to determine the B fields for which the edge and bulk states are out of equi-librium [29]. Such measurements on this sample (not shown) confirm that the voltage drop at the edge appears at fields where non-equilibrium edge states are present. Furthermore, a DC current through the sample that produces a Hall voltage on the order of $\hbar\omega_c$ is known to induce bulk-edge equilibration. We find that the voltage drop at the edge indeed disappears when an appropriate DC Hall voltage is applied to the sample, as shown in Fig. 4(d). The Hall voltage necessary to suppress the edge signal decreases from about 14 mV at $\nu = 2.5$ to 6 mV at $\nu = 6.5$.

For the voltage profile on the QH plateaus, very different behavior is expected. The bulk is nearly insulating because of the absence of extended states at $E_F$, and hence its potential is decoupled from the sample contacts. Its potential is instead set by the relative strength of the resistive coupling to the edge states on opposite sides of the sample. The coupling is determined by hopping conduction and expected to be spatially inhomogeneous, due to disorder and/or density gradients in the sample [6]. This explains the complex, position-dependent Hall profiles observed on the plateaus (Fig. 5). Similar results were obtained by Knott et al. [10] in much wider QHC samples using optical techniques. Note that these results indicate that most of the current flows in the bulk of the sample, with the local Hall current density associated with the LLs below $E_F$ distributed throughout the sample in a complex way determined by the hopping conductivity of the states at $E_F$ [30].

In summary, we have measured the local Hall voltage profile, and hence the local current distribution, within in a quantum Hall conductor. On the QH plateaus, we find that the Hall currents are primarily in the bulk and exhibit complex spatial variations. In the transition regions, we observe a combination of uniform bulk currents and localized currents at the sample edges. These results are consistent with previous experiments and theory on transport in the quantum Hall regime.

This work was supported by the NSF, NSERC, the AT&T Foundation, and the Packard Foundation. We acknowledge the Joint Services Electronics Program and the Berkeley Microlab for sample growth and fabrication.